\documentclass[fleqn,twoside]{article}
\usepackage{espcrc2}

\usepackage{graphicx}
\usepackage[figuresright]{rotating}

\begin{document}

\title{Weighing neutrinos with large-scale structure}
\author{Ofer Lahav 
\address{Department of Physics and Astronomy, University College London, 
Gower Street, London WC1E 6BT, UK}
and \O ystein Elgar\o y
\address{ Institute of Theoretical Astrophysics, University of Oslo, P.O. Box 1029, N-0315 Oslo, Norway}}

\begin{abstract}

While it is established that the effect of neutrinos on the evolution of cosmic
structure is small, the upper limits derived from large-scale structure 
could help significantly to constrain the absolute scale of the 
neutrino masses.  Current results from cosmology set an upper limit 
on the sum of the neutrino masses of $\sim 1\;{\rm eV}$, somewhat 
depending on the data sets used in the analyses and assumed priors 
on cosmological parameters.  In this review we discuss the 
effects of neutrinos on large-scale structure which make these 
limits obtainable.  We show the impact of neutrino masses on the 
matter power spectrum, the cosmic microwave background and the 
clustering amplitude. 
A summary of derived cosmological neutrino mass upper limits is given,
and we discuss future methods which will improve the mass upper limits
by an order of magnitude.



\end{abstract}

\maketitle

\section{INTRODUCTION}

The wealth of new data from  the cosmic microwave background 
(CMB) and large-scale structure (LSS) in the last few years  
indicate that we live in a flat Universe where  
$\sim 70\;\%$ of the mass-energy density is in the form of dark energy, 
with matter making up the remaining 30 \% .   
The WMAP data combined with 
other large-scale structure data \cite{wmap,sdsspap1} give impressive 
support to this picture. 
Furthermore, the baryons contribute only a fraction $f_{\rm b} 
= \Omega_{\rm b}/\Omega_{\rm m} \sim 0.15$ ($\Omega_{\rm b}$ and 
$\Omega_{\rm m}$ are, respectively, the contribution of baryons and 
of all matter to the total density in units of the critical density 
$\rho_c =  3H_0^2 / 8\pi G = 1.879\times 10^{-29}h^2\;{\rm g}\,{\rm cm}
^{-3}$, where $H_0 = 100 h \;{\rm km}\,{\rm s}^{-1}\,{\rm Mpc}^{-1}$ is 
the present value of the Hubble parameter) of this, so that 
most of the matter is dark. 
The exact nature of the dark matter in the Universe is still 
unknown.  Relic neutrinos are abundant in the 
Universe, and from the observations of oscillations of 
solar and atmospheric neutrinos we know that neutrinos have 
a mass \cite{sage,sno1,sno2,macro,gno,homestake,superk,gallex}
and will make 
up a fraction of the dark matter.  However, the oscillation experiments 
can only measure differences in the squared masses of the neutrinos, 
and not the absolute mass scale, so they cannot, at least not without 
extra assumptions,  tell us 
how much of the dark matter is in neutrinos.  
From general arguments on structure formation in the Universe we know 
that most of the dark matter has to be cold, i.e. non-relativistic 
when it decoupled 
from the thermal background.  Neutrinos with masses on the 
eV scale or below will be a hot component of the dark matter.  
If they were the dominant dark-matter component, structure in the 
Universe would have formed first at large scales, and smaller structures 
would form by fragmentation (the `top-down' scenario).  
However, the combined observational 
and theoretical knowledge about large-scale structure gives strong evidence 
for the `bottom-up' picture of structure formation, i.e. structure formed 
first at small scales.  Hence, neutrinos cannot make up 
all of the dark matter (see e.g. \cite{primack} for a review).       
Neutrino experiments give some constraints on how much of the dark 
matter can be in the form of neutrinos.
Studies of the energy spectrum in 
tritium decay \cite{mainz} provide an upper limit on the effective 
electron neutrino mass involved in this process of 
2.2 eV (95 \% confidence limit).
For the effective neutrino mass scale involved in neutrinoless 
double beta decay a range 0.1-0.9 eV has been inferred from the 
claimed detection of this process \cite{klapdor1,klapdor2}.  
If confirmed, this result would not only show that neutrinos are   
Majorana particles (i.e. their own antiparticles), but also that 
the neutrino masses are in a range where they are potentially 
detectable with cosmological probes.    

The structure of this review is as follows.  
Sections and 2 and 3  discuss  the 
effect of massive neutrinos on structure formation 
and on the CMB anisotropies.  
In section 4 we give an overview of 
recent cosmological neutrino mass limits and  in Section 5 we discuss
challenges for the future.

\section{THE EFFECT OF MASSIVE NEUTRINOS ON STRUCTURE FORMATION}

The relic abundance of neutrinos in the Universe today is 
straightforwardly found from the fact that they continue to 
follow the Fermi-Dirac distribution after freeze-out, and their 
temperature is related to the CMB temperature $T_{\rm CMB}$ today by  
$T_\nu = (4/11)^{1/3}T_{\rm CMB}$, giving 
\begin{equation}
n_\nu = \frac{6\zeta(3)}{11\pi^2}T_{\rm CMB}^3, 
\label{eq:relicabund}
\end{equation}
where $\zeta(3)\approx 1.202$, 
which gives $n_\nu \approx 112\;{\rm cm}^{-3}$ at present. 
By now, massive neutrinos will have become non-relativistic, so that 
their present contribution to the mass density can  
be found by multiplying $n_\nu$ with the total mass 
of the neutrinos $m_{\nu,\rm tot}$, giving 
\begin{equation}
\Omega_\nu h^2 = \frac{m_{\nu,\rm tot}}{94\;{\rm eV}},
\label{eq:omeganu}
\end{equation}
for $T_{\rm CMB}=2.726\;{\rm K}$.  Several effects could  
modify this simple relation.  If any of the neutrino chemical 
potentials were initially non-zero, or there were a sizable  
neutrino-antineutrino asymmetry, this would increase 
the energy density in neutrinos and give an additional contribution 
to the relativistic  energy density.  However, from Big Bang Nucleosynthesis 
(BBN) one gets a very tight limit on the electron neutrino chemical potential, 
since the electron neutrino is directly involved in the processes that 
set the neutron-to-proton ratio.  Also, within the standard three-neutrino 
framework one can extend this limit to the other flavours as well.   
Within the standard picture, 
equation (\ref{eq:relicabund}) should be accurate, and therefore 
any constraint on the cosmic mass density of neutrinos should translate 
straightforwardly into a constraint on the total neutrino mass, 
according to equation (\ref{eq:omeganu}).  
If a fourth, light `sterile' neutrino exists, sterile-active oscillations 
would modify this conclusion. 
Beacom et al. \cite{noneut}
showed that extra couplings, not yet experimentally 
excluded, of neutrinos may allow them to annihilate into 
light bosons at late times, and thus make a negligible contribution 
to the matter density today.  If so, equation (\ref{eq:omeganu}) 
is not valid, and hence neutrino mass limits derived from large-scale 
structure do not apply.  We shall assume that no such 
non-standard couplings of neutrinos exist.

\begin{figure}
\begin{center}
\includegraphics[width=84mm]{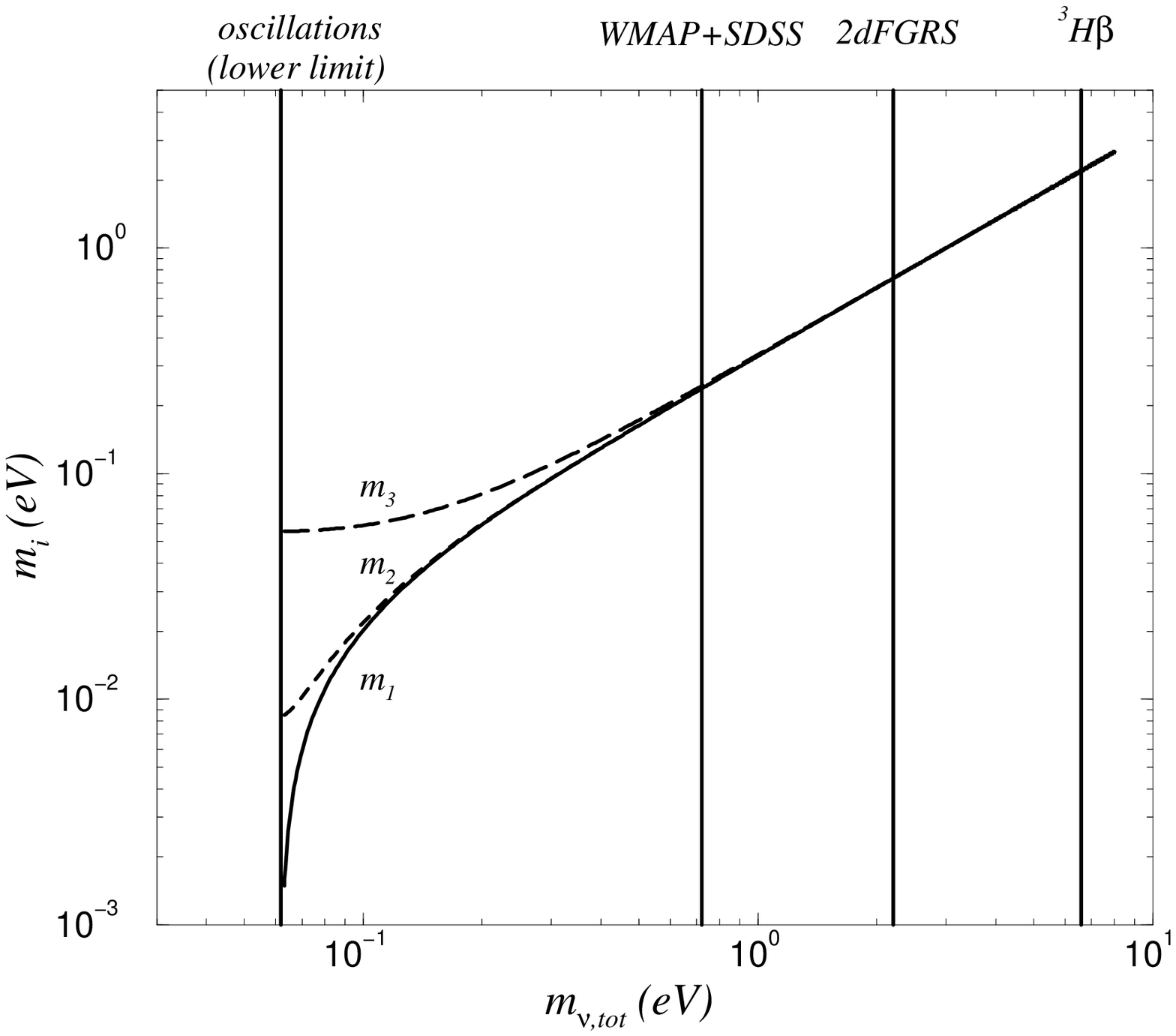}
\includegraphics[width=84mm]{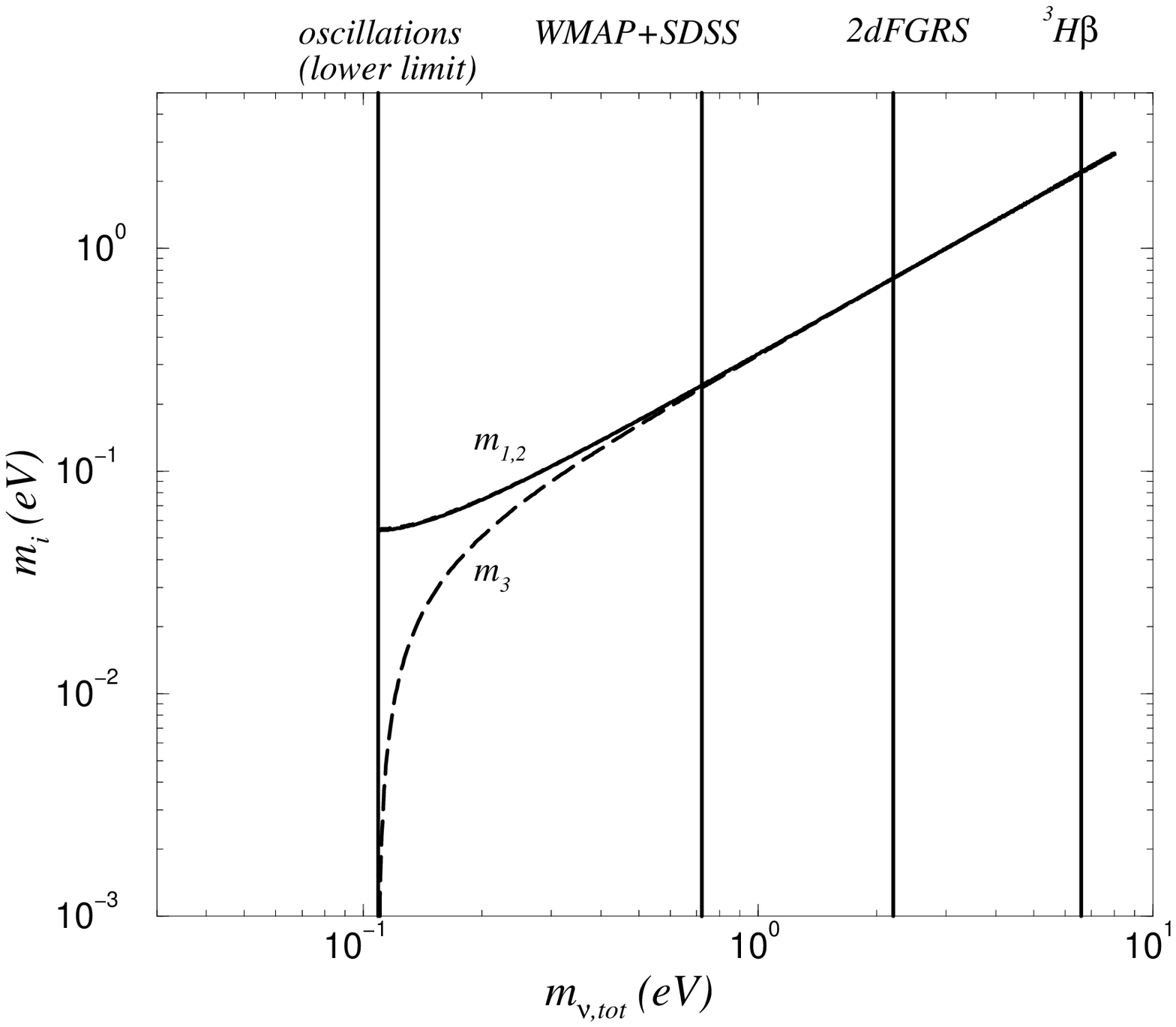}
\end{center}
\caption{Neutrino mass eigenvalues as functions of $m_{\nu,\rm tot}$ for 
the cases of normal (top panel) and inverted (bottom panel) hierarchies.  
The vertical line marked `oscillations' is the lower limit derived from 
the measured mass-squared differences. 
The vertical line marked `WMAP+SDSS' is the recent limit derived 
in \cite{seljakbias} from WMAP+SDSS, the vertical line marked 
'2dFGRS' is the limit from the 2dFGRS derived in \cite{elgar}, and 
the line marked 
`$^3{\rm H}\beta$' is the upper limit from $^3{\rm H}$ $\beta$ decay.}
\label{fig:fig1}
\end{figure}

Finally, we assume that the neutrinos are 
nearly degenerate in mass.  Current cosmological observations 
are sensitive to neutrino masses  $\sim 1\;{\rm eV}$ or greater.   
Since the mass-square differences are
small, the assumption of a degenerate mass hierarchy is therefore 
justified.  This is illustrated in Figure \ref{fig:fig1}, 
where we have plotted the mass eigenvalues $m_1,m_2,m_3$ as 
functions of $m_{\nu,\rm tot} = m_1 + m_2 + m_3$ for 
$\Delta m_{21}^2 = 7\times 10^{-5}\;{\rm eV}^2$ (solar) and 
$\Delta m_{32}^2 = 3 \times 10^{-3}\;{\rm eV}^2$ (atmospheric), 
for the cases of a normal hierarchy ($m_1 < m_2 < m_3$), and 
an inverted hierarchy ($m_3 < m_1 < m_2$).  As seen in the Figure, 
for $m_{\nu,\rm tot} > 0.4\;{\rm eV}$ the mass eigenvalues are 
essentially degenerate. 

Here we look at cosmological models 
with four components: baryons, cold dark matter, massive neutrinos, 
and a cosmological constant.  Furthermore, we restrict ourselves 
to adiabatic, linear perturbations.  The basic physics is then 
fairly simple.  
Light, massive neutrinos can  move 
unhindered out of regions below a certain limiting length scale, 
and will therefore tend to damp a 
density perturbation at a rate 
which depends on their rms velocity.    
The presence of massive neutrinos 
therefore introduces a new length scale, given by the size of the 
co-moving Jeans length when the neutrinos became non-relativistic.  
In terms of the comoving wavenumber, this is given by  
\begin{equation}
k_{\rm nr} = 0.026\left(\frac{m_{\nu}}{1\;{\rm eV}}\right)^{1/2}
\Omega_{\rm m}^{1/2}\;h\,{\rm Mpc}^{-1},
\label{eq:knr}
\end{equation}
for three equal-mass neutrinos, each with mass $m_\nu$.  
The growth of Fourier modes with 
$k > k_{\rm nr}$ will be suppressed because of neutrino free-streaming.  
The free-streaming scale varies with the cosmological epoch,
and the scale and time dependence of the power spectrum cannot 
be separated, in contrast to the situation for models with cold dark 
matter only.

The power spectrum of the matter 
fluctuations can be written as 
\begin{equation}
P_{\rm m}(k,z)= P_*(k)T^2(k,z),
\label{eq:pmatter}
\end{equation}
where $T(k,z)$ is the `transfer function', 
$P_*(k)$ is the primordial spectrum of matter fluctuations, 
commonly assumed to be a simple power law $P_*(k)=Ak^{n}$, where 
$A$ is the amplitude and the spectral index $n$ is close to 1.  
It is also common to define power spectra for 
each component, see \cite{eisenstein} for a discussion.  
Note that the transfer functions and power spectra are independent 
of the value of the cosmological constant as long as it does not 
shift the epoch of matter-radiation equality significantly.  

\begin{figure}
\begin{center}
\includegraphics[width=84mm]{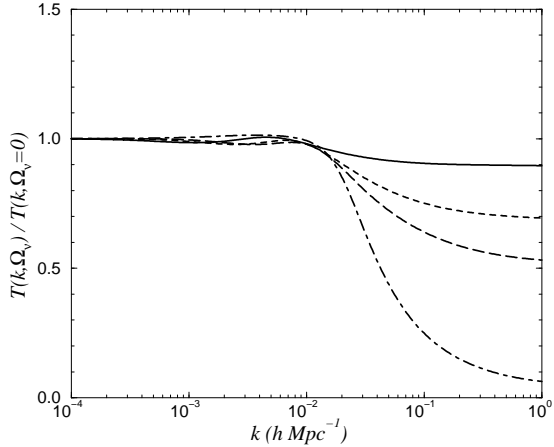}
\end{center}
\caption{Ratio of the transfer functions (at $z=0$) for various values 
of $\Omega_\nu$ to the one for $\Omega_\nu = 0$.  The other parameters 
are fixed at $\Omega_{\rm m}=0.3$, $\Omega_{\rm b}=0.04$, $h=0.7$.  
The solid line is for $m_\nu = 0.1\;{\rm eV}$, the dashed line is for 
$m_\nu=0.3\;{\rm eV}$, the long-dashed line is for $m_\nu=0.5\;{\rm eV}$, 
and the dot-dashed line corresponds to $m_\nu=2\;{\rm eV}$.}
\label{fig:transfer}
\end{figure}

The transfer function is found by solving   
the coupled fluid and Boltzmann equations for the various components.  
This can be done using one of the publicly available codes, e.g. 
CMBFAST \cite{zaldarriaga} or CAMB \cite{camb}.  
In Figure \ref{fig:transfer} we show the transfer functions for models 
with $\Omega_{\rm m}=0.3$, $\Omega_{\rm b}=0.04$, 
$h=0.7$ held constant, but with varying neutrino mass $m_\nu$.  One can 
clearly see that the small-scale suppression of power becomes more 
pronounced as the neutrino fraction $f_{\nu}\equiv \Omega_{\nu}/\Omega_
{\rm m}$ increases.

\begin{figure}
\begin{center}
\includegraphics[width=84mm]{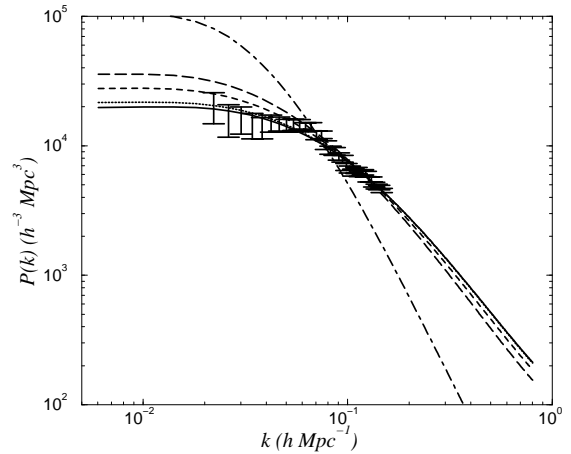}
\includegraphics[width=84mm]{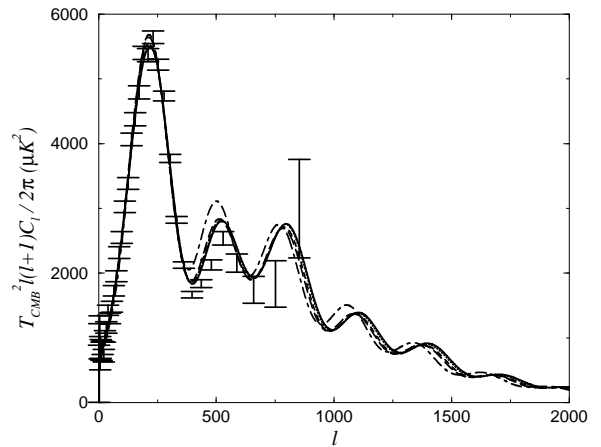}
\end{center}
\caption{{\it Top Figure:} Power spectra for $m_\nu = 0$ (full line), 
$m_\nu=0.1$ (dotted line), $m_\nu=0.3$ (dashed line), 
$m_\nu = 0.5$ (long-dashed line), and $m_\nu=3\;{\rm eV}$ (dot-dashed line).   
The other parameters 
are fixed at $\Omega_{\rm m}=0.3$, $\Omega_{\rm b}=0.04$, $h=0.7$. 
The vertical bars are the 2dFGRS power spectrum data points.  
{\it Bottom Figure:}  CMB power spectra for $m_\nu = 0$ (full line), 
$m_\nu=0.1$ (dotted line), $m_\nu=0.3$ (dashed line), 
$m_\nu = 0.5$ (long-dashed line), and $m_\nu=3\;{\rm eV}$ (dot-dashed line).   
The other parameters 
are fixed at $\Omega_{\rm m}=0.3$, $\Omega_{\rm b}=0.04$, $h=0.7$.  
The vertical bars are the WMAP power spectrum data points.}  
\label{fig:fig3}
\end{figure}

The effect is also seen in the power spectrum, as shown in 
Figure \ref{fig:fig3} (top).  Note that the power spectra shown in 
the Figure have been convolved with the 2dFGRS window function, 
as described in \cite{percival}.  
Furthermore, we have taken the possible bias of the distribution 
of galaxies with respect to that of the dark matter into 
account by leaving 
the overall amplitude of each  power spectrum as a free 
parameter to be fitted to the 2dFGRS power spectrum data (the 
vertical bars in the Figure).  For a discussion of bias in 
the context of neutrino mass limits, see \cite{olahav}.    
Because the errors on the data points are smaller at small scales, 
these points are given most weight in the fitting, and hence the power spectra 
in the Figure actually deviate more and more from each other on {\it large} 
scales as $m_\nu$ increases.  One can see from the Figure that a 
neutrino mass of $m_\nu = 0.5\;{\rm eV}$ or larger is in conflict with 
the data.  
The suppression of the power spectrum on small 
scales is roughly proportional to $f_{\nu}$: 
\begin{equation}
\frac{\Delta P_{\rm m}(k)}{P_{\rm m}(k)} \approx -8f_\nu .
\label{eq:fnusuppr}
\end{equation}  
This result can be derived from the equation of linear growth of 
density perturbations and the fact that only a fraction $(1-f_\nu)$ 
of the matter can cluster when massive neutrinos are present \cite{bond}.

\section{CONSTRAINTS FROM THE CMB ALONE}
 
Neutrino masses 
also give rise to  effects in the CMB power spectrum.  
If their masses are smaller than the temperature at 
recombination $\sim 0.3\;{\rm eV}$, their effect is very similar 
to that of massless neutrinos \cite{crotty1}.  For slightly larger 
masses, there is an enhancement of the acoustic peaks with 
respect to the massless case, as shown in Figure \ref{fig:fig3} (bottom). 
While there is some sensitivity to the neutrino mass,
note that all other parameters 
have been fixed in Figure \ref{fig:fig3} (bottom).
Analytic considerations by \cite{fukugita04} provide insight into the 
effect of the neutrinos on the CMB.
There are severe degeneracies between $m_\nu$ and other parameters 
like $n$ and $\Omega_{\rm b}h^ 2$.  The full analysis of the WMAP 
data alone in \cite{tegmarksdss} gave no upper limit on $m_\nu$.     
On the other hand  \cite{fukugita04}
have claimed an upper limit of 2.2 eV from CMB alone, in contrast 
with the conclusions of \cite{olahav} and \cite{tegmarksdss}.
The differences might be due to the assumed priors and the 
marginalisation procedures 
over other cosmological parameters.

Future CMB missions like Planck
will provide high-resolution maps of the CMB temperature and polarization 
anisotropies.  Gravitational lensing of these maps causes distortions, 
and Kaplinghat, Knox \& Song \cite{kaplinghat} have shown that this 
effect can be used to obtain very stringent limits on neutrino masses 
from the CMB alone.  For Planck, they predict a sensitivity down to 
$0.15\;{\rm eV}$, whereas a future experiment with higher resolution 
and sensitivity can possibly reach the lower bound $\sim 0.06\;{\rm eV}$ 
set by the neutrino oscillation experiments.


\section{RECENT COSMOLOGICAL NEUTRINO MASS LIMITS}

The connection between neutrino masses and cosmic structure formation 
 was realized early, but for a long time cosmologists were  
mostly interested in neutrino masses in the $\sim 10\;{\rm eV}$ range, 
since then they would be massive enough to make up all of the dark matter.  
The downfall of the top-down scenario of structure formation, 
and the fact that no evidence for neutrino masses existed before 
Super-Kamiokande detected oscillations of atmospheric neutrinos in 1998, 
makes it understandable that there was very little continuous interest 
in this sub-field.  However, the detection of neutrino oscillations 
showed that neutrinos indeed have a mass.  In an important paper 
Hu, Eisenstein \& Tegmark \cite{hutegmark} showed that one could 
obtain useful upper limits on neutrino masses from a galaxy redshift 
survey of the size and quality of the Sloan Digital Sky Survey (SDSS).  
Going down Table \ref{tab:tab1} one notes a marked improvement 
in the constraints after the 2dFGRS power spectrum became available.
After WMAP, there is a further tendency towards stronger 
upper limits, reflecting the dual role of the CMB and large-scale 
structure in constraining neutrino masses: the matter power spectrum 
is most sensitive to neutrino masses, but one needs good constraints 
on the other relevant cosmological parameters to break degeneracies 
in order to obtain low upper limits.  The limit will depend on the 
datasets and priors used in the analysis, but it seems like we 
are now converging to the precision envisaged in \cite{hutegmark}. 
The latest limit from \cite{seljakbias} uses galaxy-galaxy lensing 
to extract information about the linear bias parameter in the 
SDSS, making a direct association between the galaxy and matter 
power spectra, and hence get a stronger constraint on the neutrino 
mass than would have been possible using just the shape of the 
galaxy power spectrum.      

\begin{table*}[htb]
\begin{tabular}{l|l|l|l|l} 
\hline 
Reference & CMB & LSS  & Other& $m_{\nu,\rm tot}$ \\ 
          &     &      & data   &  limit          \\ \hline \hline
\cite{croft}    & ---   & Ly$\alpha$ & COBE norm., & 5.5 eV \\ 
                &       &            &$h=0.72\pm 0.08$,  &  \\
                &       &            & $\sigma_8=0.56\Omega_{\rm m}^{0.47}$ & \\
\cite{fukugita} & ---   & $\sigma_8$ &$\Omega_{\rm m}<0.4$, & 2.7 eV \\ 
                &       &            & $\Omega_{\rm b}h^2=0.015$,   \\
                &       &          & $h<0.8$, $n=1.0$&  \\ 
\cite{wang1} & pre-WMAP & PSCz, Ly$\alpha$ & --- & 4.2 eV \\ 
\cite{elgar}  & None & 2dFGRS & BBN, SNIa, & 2.2 eV \\ 
              &      &        &HST, $n=1.0\pm 0.1$ &  \\ 
\cite{steen1}      & pre-WMAP & 2dFGRS & --- & 2.5 eV  \\ 
\cite{lewis}  &pre-WMAP & 2dFGRS & SNIa, BBN &  0.9 eV \\ 
\cite{wmap}   & WMAP+CBI+ACBAR & 2dFGRS & Ly$\alpha$ &  0.71 eV \\ 
\cite{steen3}     & WMAP+Wang comp. & 2dFGRS & HST, SNIa & 1.01 eV \\ 
\cite{allen}    & WMAP+CBI+ACBAR  & 2dFGRS & X-ray & $0.56^{+0.30}_{-0.26}
\;{\rm eV}$ \\ 
\cite{tegmarksdss}   & WMAP            & SDSS   &  ---     & 1.7 eV \\ 
\cite{barger}    & WMAP            & 2dFGRS+SDSS & --- & 0.75 eV  \\ 
\cite{crotty1}    & WMAP+ACBAR      & 2dFGRS+SDSS &---  & 1.0 eV  \\
\cite{seljakbias} & WMAP            & SDSS       & bias & 0.54 eV \\ 
\hline \hline  
\end{tabular}
\label{tab:tab1}
\end{table*}

\begin{figure}
\begin{center}
\includegraphics[width=84mm]{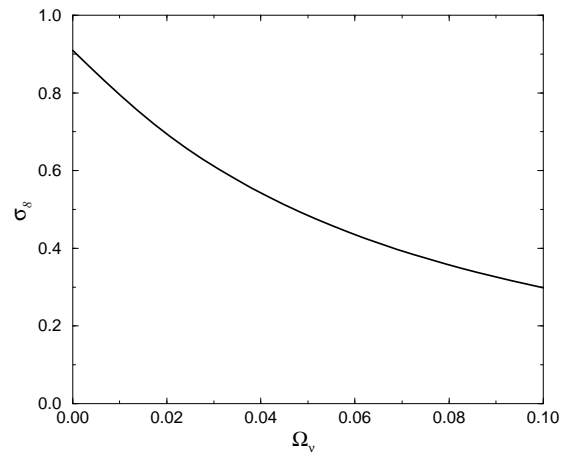}
\end{center}
\caption{The clustering amplitude $\sigma_8$ as a function of 
$\Omega_\nu$ for models with amplitude fitted to the WMAP data.}
\label{fig:fig6}
\end{figure}

Direct probes of the 
total matter distribution avoid the issue of bias and are therefore 
ideally suited for providing limits on the neutrino masses. 
Several ideas for how this can be done exist.  In \cite{fukugita} 
the normalization 
of the matter power spectrum on large scales derived from COBE 
was combined with constraints on $\sigma_8$ 
(defined as the rms mass fluctuation 
in $8h^{-1}$Mpc radius sphere), they obtained a 95\% confidence)
from cluster abundances 
and a constraint $m_{\nu,\rm tot} < 2.7\;{\rm eV}$ obtained, although 
with a fairly restricted parameter space.  However, $\sigma_8$ is 
probably one of the most debated numbers in cosmology at the moment 
\cite{wang},  
and a better understanding of systematic uncertainties connected with 
the various methods for extracting it from observations is needed 
before this method can provide useful constraints.  The potential 
of this method to push the value of the mass limit down also depends 
on the actual value of $\sigma_{\rm 8}$: the higher $\sigma_8$ turns 
out to be, the less room there will be for massive neutrinos.  
As an illustration we show in Figure \ref{fig:fig6} the value of 
$\sigma_8$ as a function of varying $\Omega_\nu$ with the 
remaining cosmological parameters fixed at their `concordance' values.  
For a given value of $m_\nu$, one fits the corresponding CMB power spectrum 
to the data.  This in turn leads to a best-fit amplitude and a prediction 
for $\sigma_8$ for the given value of $m_\nu$.  If one then has an 
independent measurement of $\sigma_8$, one can infer the value of 
$m_\nu$.   
In Figure \ref{fig:fig6} the amplitude of the power spectrum has 
been fixed by fitting to the WMAP data.  
The claimed detection of a non-zero neutrino mass in \cite{allen}  
can be seen to be due to the use of the cluster X-ray 
luminosity function to constrain $\sigma_8$, giving $\sigma_8=0.69
\pm 0.04$ for $\Omega_{\rm m}=0.3$ \cite{allen2}.  If a value 
of $\sigma_8$ at the higher end of the results reported in the 
literature is used instead, e.g. $\sigma_8=0.9$ for $\Omega_{\rm m}=0.3$ 
from \cite{nbahcall}, one gets a very tight upper limit on 
$m_\nu$, but no detection of $m_\nu > 0$.  It is clearly important 
that systematic issues related to the various methods of obtaining 
$\sigma_8$ are settled.  The evolution of cluster abundance with 
redshift may  provide  
further constraints on neutrino masses \cite{lukash}.  

Direct probes of the mass distribution such as peculiar velocities and gravitational 
lensing are also potentially important for setting constraints on the neutrino mass. 
Deep and wide weak lensing 
surveys will in the future make it possible to do weak lensing tomography 
of the matter density field \cite{hulens1,hulens2}.  By binning 
the galaxies in a deep and wide survey in redshift, one can probe the 
evolution of the gravitational potential.  However, because massive 
neutrinos and dark energy have similar effects on this evolution, 
complementary information is required in order to break this degeneracy.  
Several studies of the potential of lensing tomography to constrain 
cosmological parameters, in particular dark energy and neutrino masses, 
have been carried out, see e.g. \cite{knox} for an overview.  
Even when taking the uncertainties in the properties of dark energy 
into account, the combination of weak lensing tomography and 
high-precision CMB experiments may be sensitive to neutrino masses 
below to lower bound of 0.06 eV on the sum of the neutrino masses 
set by the current oscillation data \cite{knox}.

\section{DISCUSSION}

The dramatic increase in amount and quality of CMB and large-scale 
structure data we have seen in cosmology in the last few years 
have made it possible to derive fairly stringent limits on the 
neutrino mass scale.  With the WMAP and SDSS data, the upper limit 
has been pushed down to $\sim 1\;{\rm eV}$ for the total mass,  
assuming three massive neutrino species. 

One point to bear in mind is that all these limits assume the 
`concordance' $\Lambda{\rm CDM}$ model with adiabatic, scale-free 
primordial fluctuations.  While the wealth of cosmological data 
strongly indicate that this is the correct basic picture, one 
should keep in mind that cosmological neutrino mass limits 
are model-dependent, and that there might still be surprises.
As the suppression of the power spectrum depends on the ratio
$\Omega_{\nu}/\Omega_{\rm m}$, \cite{elgar} found that the
out-of-fashion Mixed Dark Matter (MDM) model, with $\Omega_{\nu}=0.2$,
$\Omega_{\rm m}=1 $ and no cosmological constant, fits the 2dFGRS power
spectrum well, but only for a Hubble constant $H_0 < 50\;{\rm km}\,{\rm
s}^{-1}\,{\rm Mpc}^{-1}$. 
A similar conclusion was reached in 
\cite{blanchard}, and they also found that the CMB power spectrum could be
fitted well by the same MDM model if one allows features in the
primordial power spectrum.  
Another consequence of
this is that excluding low values of the Hubble constant, e.g. with
the HST Key Project, is important in order to get a strong upper limit
on the neutrino masses.

If the future observations live up to their promise, 
the prospects for pushing the cosmological neutrino mass limit 
down towards $0.1\;{\rm eV}$ are good.  Then, 
as pointed out in 
\cite{lesgourgueslatest}, one may even  
start to see effects of the different mass hierarchies (normal or 
inverted), and thus one should take this into account when 
calculating CMB and matter power spectra.  For example, with a 
non-degenerate mass hierarchy one will get more than one free-streaming 
scale, and this will leave an imprint on the matter power spectrum.  
The coming years will see 
further comparison between the effective neutrino mass in Tritum beta decay, 
the effective Majorana neutrino mass in neutrinoless double beta decay and the 
sum of neutrino mass from Cosmology (\cite{fogli04}).
It would be a great triumph for cosmology if the neutrino mass 
hierarchy were finally revealed by the distribution of large-scale 
structures in the Universe.

\section*{Acknowledgements}
{\O}E acknowledges support from the Research Council of Norway 
through grant number 159637/V30 and 
OL thanks PPARC for a Senior Research Fellowship.


\end{document}